\documentstyle[prl,aps,twocolumn,epsfig]{revtex}
\begin{document}
\onecolumn

\draft

\title{Measurement of $G_{Ep}/G_{Mp}$ in $\vec ep\rightarrow e\vec p$ to $Q^2 = 5.6$ GeV$^2$}

\author {
O. Gayou,$^{1,7}$
K.A.~Aniol,$^{9}$
T.~Averett,$^{1}$
F.~Benmokhtar,$^{2}$
W.~Bertozzi,$^{24}$
L.~Bimbot,$^{26}$
E.J.~Brash,$^{4}$
J.R.~Calarco,$^{25}$
C.~Cavata,$^{27}$
Z.~Chai,$^{24}$
C.-C.~Chang,$^{23}$
T.~Chang,$^{15}$
J.-P.~Chen,$^{5}$
E.~Chudakov,$^{5}$
R.~De Leo,$^{16}$
S.~Dieterich,$^{2}$
R.~Endres,$^{2}$
M.B.~Epstein,$^{9}$
S.~Escoffier,$^{27}$
K.G.~Fissum,$^{22}$
H.~Fonvieille,$^{7}$
S.~Frullani,$^{18}$
J.~Gao,$^{8}$
F.~Garibaldi,$^{18}$
S.~Gilad,$^{24}$
R.~Gilman,$^{2,5}$
A.~Glamazdin,$^{21}$
C.~Glashausser,$^{2}$
J.~Gomez,$^{5}$
V.~Gorbenko,$^{21}$ 
J.-O.~Hansen,$^{5}$
D.W.~Higinbotham,$^{24,}$\footnotemark[1]
G.M.~Huber,$^{4}$
M.~Iodice,$^{17}$
C.W.~de~Jager,$^{5}$
X.~Jiang,$^{2}$
M.K.~Jones,$^{5}$
J.J.~Kelly,$^{23}$
M.~Khandaker,$^{3}$
A.~Kozlov,$^{4}$
K.M.~Kramer,$^{1}$
G.~Kumbartzki,$^{2}$
J.J.~LeRose,$^{5}$
D.~Lhuillier,$^{27}$
R.A.~Lindgren,$^{26}$
N.~Liyanage,$^{5}$
G.J.~Lolos,$^{4}$
D.J.~Margaziotis,$^{9}$
F.~Marie,$^{27}$
P.~Markowitz,$^{12}$
K.~McCormick,$^{20}$
R.~Michaels,$^{5}$
B.D.~Milbrath,$^{11}$
S.K.~Nanda,$^{5}$
D.~Neyret,$^{27}$
Z.~Papandreou,$^{4}$
L.~Pentchev,$^{1,}$\footnotemark[2]
C.F.~Perdrisat,$^{1}$
N.M.~Piskunov,$^{19}$
V.~Punjabi,$^{3}$
T.~Pussieux,$^{27}$
G.~Qu\'em\'ener,$^{14}$
R.D.~Ransome,$^{2}$
B.A.~Raue,$^{12}$
R.~Roch\'e,$^{13}$
M.~Rvachev,$^{24}$
A.~Saha,$^{5}$
C.~Salgado,$^{3}$
S.~\v{S}irca,$^{24}$
I.~Sitnik,$^{19}$
S.~Strauch,$^{2,}$\footnotemark[3]
L.~Todor,$^{10}$
E.~Tomasi-Gustafsson,$^{27}$
G.M.~Urciuoli,$^{18}$
H.~ Voskanyan,$^{29}$
K.~Wijesooriya,$^{6}$
B.B.~Wojtsekhowski,$^{5}$	
X.~Zheng,$^{24}$
L.~Zhu$^{24}$			
\vspace{0.1in}
}
\footnotetext[1]{Currently at Thomas Jefferson National Accelerator Facility}
\footnotetext[2]{On leave of absence from Institute for Nuclear Research and Nuclear Energy, Sofia, Bulgaria}
\footnotetext[3]{Currently at George Washington University, Washington, DC 20052}

\address{
{(for the Jefferson Lab Hall A Collaboration)}\\
\vspace{0.1in}
$^{1}$College of William and Mary, Williamsburg, VA 23187\\
$^{2}$Rutgers, The State University of New Jersey,  Piscataway, NJ 08855\\
$^{3}$Norfolk State University, Norfolk, VA 23504\\
$^{4}$University of Regina, Regina, SK S4S OA2, Canada\\
$^{5}$Thomas Jefferson National Accelerator Facility, Newport News, VA 23606\\
$^{6}$Argonne National Laboratory, Argonne, IL 60439 \\
$^{7}$Universit\'{e} Blaise Pascal/CNRS-IN2P3, F-63177 Aubi\`{e}re, France\\ 
$^{8}$California Institute of Technology, Pasadena, CA 91125\\
$^{9}$California State University, Los Angeles, CA 90032\\
$^{10}$Carnegie-Mellon University, Pittsburgh, PA 15213\\
$^{11}$Eastern Kentucky University, Richmond, KY 40475\\
$^{12}$Florida International University, Miami, FL 33199\\
$^{13}$Florida State University, Tallahassee, FL 32306\\
$^{14}$Institut des Sciences Nucl\'{e}aires, CNRS-IN2P3, F-38026 Grenoble, 
France\\
$^{15}$University of Illinois, Urbana-Champagne, IL 61801\\
$^{16}$INFN, Sezione di Bari and University of Bari, 70126 Bari, Italy\\
$^{17}$INFN, Sezione di Roma-III, 00146 Roma, Italy\\
$^{18}$INFN, Sezione Sanit\`{a} and Istituto Superiore di Sanit\`{a}, 
00161 Rome, Italy\\
$^{19}$JINR-LHE, 141980 Dubna, Moscow Region, Russian Federation\\
$^{20}$Kent State University, Kent, OH 44242\\
$^{21}$Kharkov Institute of Physics and Technology, Kharkov 61108, Ukraine\\
$^{22}$University of Lund, PO Box 118, S-221 00 Lund, Sweden\\
$^{23}$University of Maryland, College Park, MD 20742\\
$^{24}$Massachusetts Institute of Technology, Cambridge, MA 02139\\
$^{25}$University of New Hampshire, Durham, NH 03824\\
$^{26}$Institut de Physique Nucl\'{e}aire, F-91406 Orsay, France\\
$^{27}$DAPNIA/SPhN CEA/Saclay, F-91191 Gif-sur-Yvette, France\\
$^{28}$University of Virginia, Charlottesville, VA 22901\\
$^{29}$Yerevan Physics Institute, Yerevan 375036, Armenia\\
}

\date{\today}

\maketitle

\begin{abstract}
The ratio of the electric and magnetic form factors of the proton, 
$G_{Ep}/G_{Mp}$, was measured at the Thomas Jefferson National Accelerator Facility
(JLab) using the recoil polarization technique. 
The ratio of the form factors is directly
proportional to the ratio of the transverse to longitudinal components
of the polarization of the recoil proton in the elastic $\vec ep \rightarrow
e\vec p$  reaction. The new data presented in this article span the range 
$3.5 < Q^2 < 5.6$~GeV$^2$ and are well described by a linear $Q^2$ fit. Also,
the ratio $QF_{2p}/F_{1p}$ reaches a constant value above $Q^2=2$~GeV$^2$.
\end{abstract}



\twocolumn




The nucleon electromagnetic form factors are a key ingredient to 
describe its internal structure, and eventually understand the strong
interaction. 
Experimental values for the
proton have been obtained over the last 50 years 
via electron-proton scattering, often using the Rosenbluth separation 
technique~\cite{rosenbluth}. They 
show that the magnetic form factor, $G_{Mp}$, follows approximately a 
dipole form factor $G_D = \left (1+{Q^2}/{0.71 ({\rm GeV^2})} \right )^{-2}$ 
where $Q^2$ is the four-momentum
transfer squared~\cite{litt,berger,price,bartel,sill,walker,andivahis}. 
However, measuring the charge form factor $G_{Ep}$ 
by Rosenbluth separation
becomes difficult for $Q^2>1$~GeV$^2$, because the charge scattering 
contributes only little to the differential cross section.
Extending the measurement of the form factors to larger $Q^2$ is important,
for example to test the perturbative QCD (pQCD) scaling predictions for the
Dirac and Pauli form factors $F_{1p}$ and $F_{2p}$~\cite{brodsky}.
The recoil polarization method, proposed in the 1970's~\cite{akharn}, has been 
established as the most effective available technique for measuring the ratio 
$G_{Ep}/G_{Mp}$ at large $Q^2$~\cite{jones,milbrath,dieterich,sonja}. The
results of Ref.~\cite{jones} showed a surprising, roughly linear, decrease 
of this ratio as a function of $Q^2$ up to $3.5$~GeV$^2$. In a 
non-relativistic approach, this faster 
decrease of $G_{Ep}$ can be interpretated as  
confinement of the charge distribution in the Breit frame to a larger region 
of space than the magnetism distribution.

In the one-photon exchange approximation
for elastic $ep$ scattering, a longitudinally polarized electron
beam transfers its polarization to the recoil proton with two non-zero
components, $P_t$, perpendicular to, and $P_{\ell}$, 
parallel to, the proton 
momentum in the scattering plane. $P_t$ and $P_{\ell}$ 
are proportional to $G_{Ep}G_{Mp}$ and $G_{Mp}^2$, respectively, so that 
the ratio of the form factors follows directly from the  
simultaneous measurements of these two polarization components~\cite{akharn}:
\begin{equation}
\frac{G_{Ep}}{G_{Mp}}=-\frac{P_t}{P_{\ell}}\frac{\left (E_e+E_{e^{\prime}} 
\right )}{2m}\tan \frac{\theta _e}{2}
\label{ffratio}
\end{equation} 
Here $m$ is the proton mass, 
$\theta _e$ is the lab scattering angle, and $E_e$ and 
$E_{e^{\prime}}$ are the incident and scattered energies of the electron.


We present the results of new measurements of the ratio $\mu _pG_{Ep}/G_{Mp}$,
where $\mu _p$ is the magnetic moment of the proton,
up to $Q^2=5.6$~GeV$^2$ performed in Hall A at Jefferson Lab.
A polarized electron beam from the Continuous Electron Beam 
Accelerator was
scattered on a 15~cm-long circulating liquid hydrogen target. 
A strained GaAs crystal excited by circularly polarized
laser light produced the polarized electron beam, with an average current of 
$40$~$\mu$A. A typical longitudinal beam polarization at the target of 
$\sim 0.70$ was measured with
both a M\o ller polarimeter ~\cite{moller} (with an uncertainty of $\sim$3\%) 
and a Compton polarimeter~\cite{compton} (with 
an uncertainty of $\sim$1.4\%~\cite{steph}). The 
helicity of the beam was flipped pseudo-randomly at $30$~Hz. 


Recoil protons were detected in the left high resolution spectrometer 
(HRS)~\cite{hallA}. The HRS has a central bend angle of 
$45^\circ$, and 
accepts a maximum central momentum of $4$~GeV/c with a $6.5$~msr
acceptance; it has a $\pm 5\%$ momentum acceptance and a $<2\times 10^{-4}$ 
momentum resolution. Two vertical drift chambers located at the focal plane, 
along with the knowledge of the optics of the three quadrupoles and the dipole 
of the 
HRS, allow precise position and angle measurements of the proton 
trajectory at the target. As the data acquisition was triggered by a single
proton in the HRS, we also detected the scattered electron in order to
isolate elastic $ep$ scattering events and reject the
significant background in the spectrometer, mostly from pion 
electroproduction.  
The polarization transfer in this 
reaction can be different in magnitude and sign from the polarization
transfer in elastic scattering. 

\begin{center}
\begin{figure}[b!]
\centerline{\epsfxsize=200pt \epsfysize=140pt \epsfbox{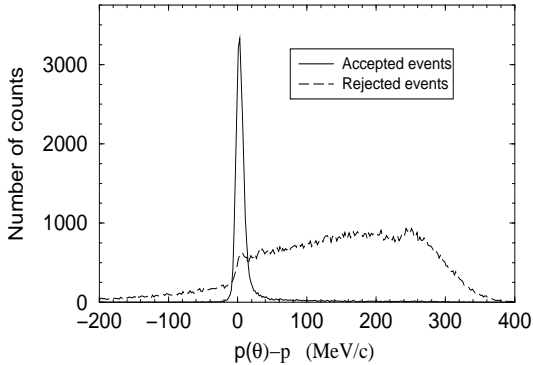}}
\caption{Selection of elastic events by the calorimeter. The histograms show 
the spectrum of accepted and rejected events
of the momentum difference between the proton momentum expected from 
its 
reconstructed scattering angle $\theta$ and elastic kinematics calculation, 
$p(\theta )$, and its 
momentum measured by the HRS, $p$.}
\label{fig:pcpo}
\end{figure}
\end{center}

For the measurement at $Q^2=3.5$~GeV$^2$, the electron was detected in
the second (right) HRS, 
and the trigger was a coincidence between an electron and a proton,
as described in Ref.~\cite{jones}.  
For the measurements at higher $Q^2$, at a fixed beam energy of 
$4.6$~GeV, 
the electron was scattered at a larger angle than the proton, 
and thus defined
the rate of the reaction. To maximize the number of elastic events 
selected, the electron was detected in a calorimeter with a large solid angle.
The $1.35\times 2.55$~m$^2$ calorimeter was assembled with blocks of 
lead-glass with a cross-sectional area of $15\times 15$~cm$^2$ each, in 
$9$~columns 
and $17$~rows. The use of lead-glass, which produces \v{C}erenkov light, 
provides good pion background suppression.
At each $Q^2$, the calorimeter was located at a distance from the target
where the electron solid angle matched the proton
HRS acceptance according to the Jacobian of the reaction. This distance ranges
from 9~m at $Q^2=5.6$~GeV$^2$ to 17~m at $Q^2=4.0$~GeV$^2$.
The trigger was defined by a proton in the HRS, signaled by a coincidence of 
two planes of scintillators in the focal plane.
For each single proton event in the left HRS, the ADC and TDC 
information from the calorimeter was read out for all blocks, and elastic 
events were selected by applying software cuts
to the calorimeter data.
Our analysis showed that the calorimeter registered an ADC
signal in about ten blocks 
for each trigger. A tight coincidence time cut was applied
to ensure that the particle detected
in the calorimeter came from the same reaction that produced the proton.
This considerably reduced that part of the pion electroproduction background 
for which the 
scattered electron and 
the photons from decay of the $\pi ^0$ were 
mainly not in the acceptance of the calorimeter.
A cut was applied to the angular 
correlation between the proton and the electron to reject 
events where the pion production products happened to be in the 
acceptance. The remaining background represents less than 1\% of the 
accepted events, and is taken into account in the polarization analysis,
by measuring the polarization of the rejected events. 
The small bump in the elastic region of the rejected events in 
Fig.~\ref{fig:pcpo} shows that about $5\%$ of elastic events are rejected, 
because of missing lead-glass blocks in the calorimeter.

\begin{center}
\begin{figure}[b]
\centerline{\epsfxsize=200pt \epsfysize=190pt \epsfbox{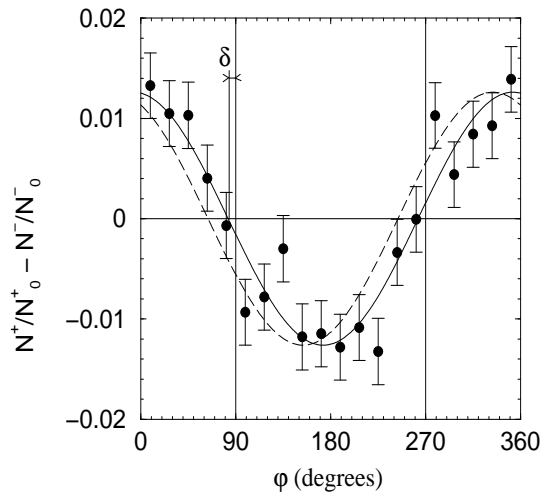}}
\caption{Difference distribution for positive and 
negative electron beam helicity, for $Q^2=5.6$~GeV$^2$. See text for details.} 
\label{fig:asym}
\end{figure}
\end{center}

The recoil proton polarization was measured by the focal plane polarimeter
(FPP) located behind the focal plane of the left HRS~\cite{FPP}. The FPP 
determines the two polarization
components perpendicular to the momentum, 
$P^{fpp}_t$ and $P^{fpp}_n$,
by measuring asymmetries in 
the azimuthal angular distribution after scattering the proton in an 
analyzer. 
To improve the figure of merit,
the usual graphite analyzer was replaced by
polyethylene, $60$~cm thick at $Q^2 = 3.5$~GeV$^2$ and 
$100$~cm thick for the other kinematics. 
The angular distribution is measured by detecting the trajectory
of the proton in two 
sets of two straw chambers,
one before and one after the scattering in the analyzer; 
the distribution is given by:
\begin{eqnarray} \nonumber
N(\vartheta ,\varphi ) = N_0(\vartheta)\{1+[A_y(\vartheta)P_t^{fpp} + a_{in}]\sin \varphi \\
-[A_y(\vartheta)P_n^{fpp} + b_{in}]\cos \varphi \}
\label{eq:azimdistr}
\end{eqnarray} 
where $N_0(\vartheta)$ is the number of 
protons scattered in the polarimeter to a polar angle $\vartheta$,  
$\varphi$ is the 
azimuthal angle after scattering, and $A_y(\vartheta)$ is the analyzing
power; $a_{in}$ and $b_{in}$ are instrumental asymmetries. 
Such a distribution was measured for the two states of the electron
beam helicity, positive and negative. The difference
in the beam polarization for these two helicity states was
compatible with zero at the 0.3\% level~\cite{steph}. 
The difference between these two distributions $N^+/N^+_0-N^-/N^-_0$ 
cancels the instrumental asymmetries to first order. It also
gives us access to the transferred, helicity-dependent polarization, which
is the quantity of interest. The induced, 
helicity-independent
polarization is zero in the 
case of elastic scattering from the proton. 
Figure~\ref{fig:asym} shows this difference distribution, fitted (solid line) 
with a cosine function $C\cos (\varphi + \delta)$, where the amplitude $C$ is 
$\sqrt{{(P_{n}^{fpp})}^2+{(P_{t}^{fpp})}^2}$ and the phase shift $\delta$ is 
such that $\tan \delta = P_t^{fpp}/P_n^{fpp}$. Since $P_t^{fpp}$ is 
related to the interference term $G_{Ep}G_{Mp}$, this phase shift is 
a measure of $G_{Ep}$. 
The dashed line
represents what the distribution would look like if $\mu _pG_{Ep}/G_{Mp}=1$.
The vertical lines
at $\varphi =90^{\circ}$ and $\varphi =270^{\circ}$ emphasize the phase shift 
$\delta$.


The proton spin precesses through the magnetic fields of the
HRS. The polarization vector at the analyzer of the FPP, $\mathbf{P^{fpp}}$, 
is related to the 
polarization vector at the target, $\mathbf{P}$, by the spin transfer matrix 
$\mathbf{S}$:
$\mathbf{P^{fpp}=S \times P}$. Because protons with different 
angles and interaction points at the target see different magnetic fields in 
the HRS, the 
matrix elements $S_{ij}$ must be calculated for each event 
from the reconstructed target coordinates.
The matrix elements were determined using a model of the HRS based on
optics studies and using the differential algebra-based code COSY~\cite{bertz}. 
 
The polarization components $hA_yP_t$ and $hA_yP_{\ell}$ are obtained by
maximizing the likelihood function~\cite{steffen} $L(P_t,P_{\ell})$ defined as
\begin{eqnarray} \nonumber
L(P_t,P_{\ell}) = \prod_{i=1}^{N_p} \{1\pm A_y(\vartheta _i)(S_{tt,i}hP_t+S_{t\ell ,i}hP_{\ell})\sin \varphi_{i} \\
\mp  A_y(\vartheta _i)(S_{nt,i}hP_t+S_{n\ell,i}hP_{\ell})\cos \varphi_{i} \}
\label{maxlike},
\end{eqnarray} where the product runs over all events, $N_p$,
$\pm$ stands for the sign of the beam helicity and $h$ is the beam
polarization. The analyzing power and beam helicity eventually 
cancel in forming the ratio $hA_yP_t/hA_yP_{\ell}$.

\begin{center}
\begin{figure}[b]
\centerline{\epsfxsize=250pt \epsfysize=220pt \epsfbox{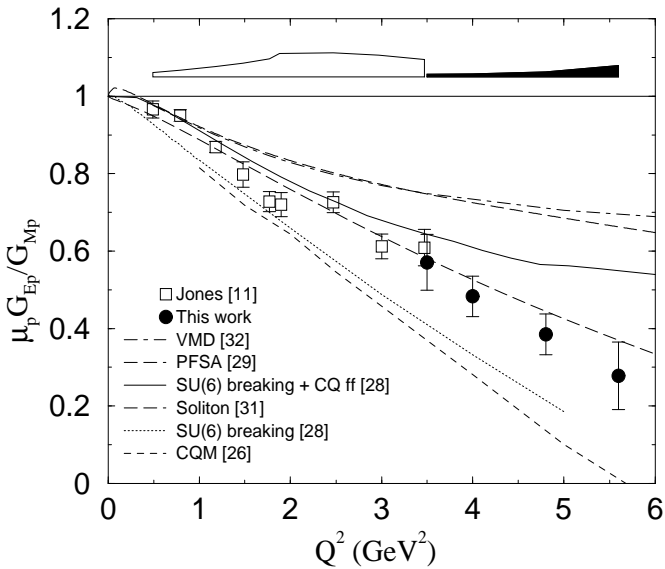}}
\caption{The ratio $\mu_p G_{Ep}/G_{Mp}$ from this experiment and
Jones {\it et al.} (Ref.~[11]), compared with theoretical calculations. 
Systematic 
errors for both experiments are shown as a band at the top of the figure.} 
\label{fig:results}
\end{figure}
\end{center}

The new results for the ratio $\mu _pG_{Ep}/G_{Mp}$ are presented in 
Fig.~\ref{fig:results}, with statistical error bars, together with the data
of Ref.~\cite{jones}.
The systematic errors are represented by the bands at the top.
The new data are tabulated in 
Table~\ref{table:results}, with their statistical and systematic errors. 
The main sources of systematic errors are related to the spin precession. 
Those can be divided into three parts. 
Our analysis shows that the major part is the error associated
with the uncertainty in the total bending angle in the non-dispersive
plane of the spectrometer, due to 
misalignment of the magnetic elements of the spectrometer.
A careful study of this misalignment has been done recently
in Hall~A~\cite{lubomir}, reducing the systematic error compared to
Ref.~\cite{jones} at $Q^2=3.5$~GeV$^2$ by a factor of six. The 
other sources of
error in the precession are related to uncertainties in the dipole fringe 
field model, and to the bending angle in the dispersive plane. 
Systematic errors associated with proton momentum, electron beam energy and 
electron scattering angle give smaller contributions.
No radiative corrections have been 
applied to the ratio, as no full calculation of polarization 
observables for $ep$ scattering exists. 
Afanasev {\it et al.}\cite{afanasev1} have calculated the 
single photon emission corrections to the two polarization observables in 
hadronic variables.
The two corrections are of the same sign, negative, and are each of the order 
of 1\%; thus they 
largely cancel when one takes the ratio. Other contributions due to two 
photon-exchange, virtual Compton scattering and interference terms
are expected to be at the percent level~\cite{afanasev2}.

A straight line fit has been applied to the ratio $\mu _pG_{Ep}/G_{Mp}$ in the 
range $0.5<Q^2<5.6$~GeV$^2$:
\begin{equation}
\mu _p\frac{G_{Ep}}{G_{Mp}}=1-0.13(Q^2-0.04)
\label{eq:linfit}
\end{equation}
Using this $Q^2$-dependence as a constraint on $G_{Ep}$, the Rosenbluth 
separation data have been reanalyzed. This brings a correction of the order of 1.5 to 3\% to the magnetic form
factor~\cite{brash}.


Also shown in Fig.~\ref{fig:results} are the results of some theoretical 
calculations which discuss possible interpretations of a decrease of the ratio
$\mu _pG_{Ep}/G_{Mp}$. 
Several authors have studied different effects within the framework of the 
constituent quark model (CQM); 
all emphasize the necessity of both kinematic
and dynamic relativistic corrections. 
Franck, Jennings and Miller~\cite{miller},
in their study of nuclear medium effects on nucleon electromagnetic form 
factors, used 
Schlumpf's light-front wave function in an early relativistic 
CQM~\cite{schlumpf} 
to compute the free proton elastic form factors (dashed curve). 
Based on the data of Ref.~\cite{jones},
Cardarelli and
Simula~\cite{card} show that a suppression of the ratio can be expected 
in the CQM, if the relativistic effects generated by the 
SU(6) symmetry breaking caused by the Melosh rotations of the constituent spins
are taken into account. Their prediction is shown using point-like quark 
constituents (dotted curve) and constituent quark form factors (solid curve).
Wagenbrunn {\it et al.}~\cite{wagen} (thin long-dashed curve) reach
a reasonable agreement with all electroweak nucleon form factors in 
their point-form spectator approximation (PFSA) prediction of the Goldstone
boson exchange CQM~\cite{gloz}.
Other types of models try to describe the dynamic features of the nucleon. 
Holzwarth~\cite{holzwarth} (thick long-dashed curve) uses a relativistic 
chiral soliton model, which gives remarkable 
agreement with the data. Lomon~\cite{lomon} 
used the world data, including Ref.~\cite{jones}, to perform a fit within the 
Vector Meson Dominance (VMD) 
model, where the $\rho$ meson contribution is determined by dispersion
relations (dot-dashed curve). It is worthwhile to note that while some
models can reproduce the observed behavior of $\mu _pG_{Ep}/G_{Mp}$, they are
all based on effective theories and have parameters that can be adjusted to fit
the data. No model so far can accurately describe all form factors of the 
nucleon,
as is necessary to fully understand the strong interaction.

\begin{center}
\begin{figure}[b]
\centerline{\epsfxsize=200pt \epsfysize=230pt \epsfbox{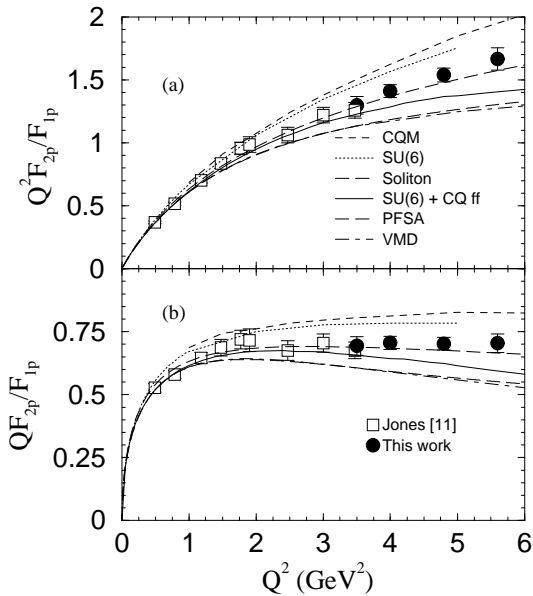}}
\caption{Same legend as Fig. \ref{fig:results}, for (a) $Q^2F_{2p}/F_{1p}$ and (b) $QF_{2p}/F_{1p}$.} 
\label{fig:ff}
\end{figure}
\end{center}

The result can also be expressed in terms of the non spin-flip Dirac form 
factor $F_{1p}$, and spin-flip Pauli form factor $F_{2p}$, given by:
\begin{eqnarray} 
F_{1p} = \frac{G_{Ep}+\tau G_{Mp}}{1+\tau }    ;     
F_{2p} = \frac{G_{Mp}-G_{Ep}}{\kappa _p\left (1+\tau \right )} 
\label{eq:pauli}
\end{eqnarray} where $\kappa _p$ is the anomalous magnetic moment of 
the proton, and $\tau=Q^2/4m^2$.
The ratio $F_{2p}/F_{1p}$ directly follows from $G_{Ep}/G_{Mp}$.
In Fig.~\ref{fig:ff}a, the results are compared with the pQCD 
predictions~\cite{brodsky} that the asymptotic behavior of the form 
factors is ${F_{1p} \propto \frac{1}{Q^4}}$ and 
${F_{2p}\propto \frac{1}{Q^6}}$, 
so that 
${Q^2 \frac{F_{2p}}{F_{1p}}}$ would reach a constant value at high enough 
$Q^2$. The 
data clearly indicate that this asymptotic regime has not been reached yet. 
Based on the results of Ref.~\cite{jones}, 
Ralston
{\it et al.}~\cite{ralston} postulated a different scaling behavior, 
where $F_{2p}/F_{1p}$ goes as $1/\sqrt{Q^2}$ instead of $1/Q^2$, arguing
that it
corresponds to the pQCD expectation if one takes into account contributions to 
the proton quark wave-function from states with non-zero orbital angular 
momentum. The ratio
$\sqrt{Q^2}\frac{F_{2p}}{F_{1p}}$ is shown on Fig.~\ref{fig:ff}b; a constant
value is clearly reached starting at $Q^2\sim 2$~GeV$^2$.

In conclusion, we have measured  
$G_{Ep}/G_{Mp}$ by polarization transfer to $Q^2=5.6$~GeV$^2$. The ratio
obtained in this experiment continues to decrease, as observed first 
in Ref.~\cite{jones}. 
Extrapolation of the linear trend indicates that
the electric form factor would cross zero at $Q^2 \sim 7.7$~GeV$^2$.
This result also reveals a flattening of the ratio $QF_{2p}/F_{1p}$ starting
at $Q^2\sim 2$~GeV$^2$. 
A measurement of $G_{Ep}/G_{Mp}$ to yet higher $Q^2$ is planned in the near 
future~\cite{prop}.

The collaboration thanks the Hall A technical staff and the Jefferson Lab 
Accelerator Division for their outstanding support during the experiment.
The Southern Universities Research Association (SURA) operates the Thomas
Jefferson National Accelerator Facility for the United States Department
of Energy under contract DE-AC05-84ER40150.
This work was also supported by the U.S. National Science Foundation and 
Department of Energy, the 
Italian Istituto Nazionale di Fisica Nucleare 
(INFN), the French Commissariat \`a l'Energie Atomique (CEA) and Centre 
National de 
la Recherche Scientifique (CNRS-IN2P3), the Natural Sciences and Engineering
Research Council of Canada (NSERC), and the EEC grant
INTAS 99-00125 for the Kharkov Institute of Physics and 
Technology, and CRDF UP2-2271.

\begin{center}
\begin{table}[p!]
\caption[]{The ratio $\mu_p G_{Ep}/G_{Mp}$ with  statistical uncertainty
(1$\sigma$) $\Delta _{stat}$, and systematic uncertainty $\Delta _{syst}$. 
$\langle Q^2\rangle $ is the value of $Q^2$ weighted-averaged over the 
acceptance, and 
$\Delta Q^2$ is the $Q^2$ acceptance (1$\sigma$).}
\begin{tabular}{ccccc}
$\langle Q^2\rangle \pm \Delta Q^2$ (GeV$^2$) & $\mu_p G_{Ep}/G_{Mp}$ & $\Delta _{stat}$  & $\Delta_{sys}$            \\ \hline  
3.50$\pm$0.23 & 0.571 & 0.072 & 0.007   \\
3.97$\pm$0.26 & 0.483 & 0.052 & 0.008   \\
4.75$\pm$0.30 & 0.385 & 0.053 & 0.011   \\
5.54$\pm$0.34 & 0.278 & 0.087 & 0.029   \\
\end{tabular}
\label{table:results}
\end{table}
\end{center}

\end{document}